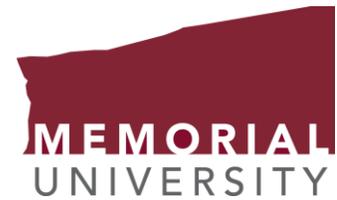

# ANALYTiC: Understanding Decision Boundaries and Dimensionality Reduction in Machine Learning

# Haidri, Salman

Department of Computer Science

Memorial University of Newfoundland and Labrador

**Supervised by Dr. Amilcar Soares**

A dissertation submitted to the Department of Computer Science in partial fulfillment of the requirements for the degree of Bachelor of Science (Honours) in Computer Science.

December 2023

# Abstract

The advent of compact, handheld devices has given us a pool of tracked movement data that could be used to infer trends and patterns that can be made to use. With this flooding of various trajectory data of animals, humans, vehicles, etc., the idea of ANALYTiC originated, using active learning to infer semantic annotations from the trajectories by learning from sets of labeled data. This study explores the application of dimensionality reduction and decision boundaries in combination with the already present active learning, highlighting patterns and clusters in data. We test these features with three different trajectory datasets with objective of exploiting the the already labeled data and enhance their interpretability. Our experimental analysis exemplifies the potential of these combined methodologies in improving the efficiency and accuracy of trajectory labeling. This study serves as a stepping-stone towards the broader integration of machine learning and visual methods in context of movement data analysis.

To Hamdan

# Acknowledgements

I would like to extend my deepest gratitude towards Dr. Amilcar Soares, without whose invaluable guidance and support this paper was not possible. His mentorship has been instrumental, and I sincerely appreciate the opportunities he's given me to explore, grow, and learn.

A huge shout-out to my friends, especially Yaksh for being an amazing friend and supporting me constantly throughout this journey.

And to my family, your support means everything to me.

# Table of contents

# List of figures

# Chapter 1

# Introduction

Positioning devices, ranging from smartphones to GPS-enabled cameras and sensors, are rapidly increasing, generating vast volumes of spatio-temporal data. This trend is driven by these devices' smaller and more affordable nature, enabling tracking of various objects such as vehicles, vessels, animals, and humans. This has led to an explosion in the volume of spatio-temporal data, necessitating specialized techniques for their effective interpretation. Trajectory data records the change in the position of an object relative to time. There are many tasks necessary to properly work with trajectory data in a data mining setup, including: (i) data fusion [31, 18]; (ii) compression [7, 21];(iii) segmentation [33, 17, 6]; (iv) classification [9, 5]; (v) clustering [7, 37]; and (vi) outlier detection [8, 4, 1].

Semantic enrichment of movement data, resulting in what we refer to as semantic or annotated trajectories, is becoming more critical as we move forward. By supplementing geometric spatio-temporal data with semantic information, these annotated trajectories provide valuable insights across various domains such as tourism, cultural heritage, traffic management, animal behavior studies, and vehicle tracking [26]. However, the lack of methods to expose the semantic dimensions of movement data, especially those that can infer trajectory labels in large datasets, remains a significant challenge [23].

Machine learning offers a promising solution for this challenge, using labeled training sets to create models or classifiers for annotating trajectories. However, obtaining these labels often requires manual annotation, which is labor-intensive and time-consuming. The challenge lies in automating trajectory annotation by analyzing intrinsic features to minimize manual intervention. Nevertheless, high performance often requires a substantial training set, implying a significant commitment from domain experts to provide sufficient examples for the classifier. Consequently, these questions were proposed in [32] to address the research questions below: (i) Can we devise an effective machine-learning method for automatic trajectory classification, reducing the demand for human-labeled trajectories?; (ii) Is this proposed machine learning method suitable and efficient for trajectory data?; (iii) How can we facilitate user involvement in trajectory labeling? A series of empirical evaluation experiments concluded that active learning effectively reduces the trajectories datasets to be labeled [32]. The ANALYTiC (AN Active Learning sYstem Trajectory Classi cation) tool represents an innovative approach to semantic enrichment in trajectory data analysis, leveraging active learning techniques to reduce the need for extensive human-annotated trajectory data [29]. However, despite the innovation, the initial version of the tool had limitations regarding high-dimensional data visualization.

Overcoming these challenges led us to three further research questions: (i) Can we employ dimensionality reduction techniques to manage complexity and enhance computational efficiency?(ii) How can the interpretation and visualization of decision boundaries contribute to understanding model performance and the nature of the classification process in high-dimensional trajectory data analysis?

To address these research questions and challenges, this work introduces the next iteration of ANALYTiC (AN Active Learning sYstem Trajectory Classification), a tool that brings forward three key enhancements. Firstly, the introduction of dimensionality reduction techniques, such as Principal Component Analysis (PCA). This powerful technique condenses the data by transforming the original high-dimensional data into a lower-dimensional space, effectively concentrating on the most informative features. As such, it reduces computational complexity and resources, thereby making the analysis process more efficient and feasible [16].

The integration of decision boundary visualization was another crucial enhancement. This allows users to see and comprehend the decisions made by the machine learning models more intuitively. Decision boundaries demonstrate the areas of input space assigned to different classes, enhancing the overall transparency of the classification process. Users can better understand the reasoning behind the model's classifications by visually representing classifier behavior, leading to more insightful decision-making [25].

Together, these enhancements have significantly upgraded the capabilities of the ANALYTiC tool, improving its handling of high-dimensional data, increasing its transparency, and enhancing its overall performance in trajectory data analysis. Secondly, the new version of ANALYTiC, a visually interactive web-based annotation tool, is designed to aid users in the active learning trajectory annotation task. Improved visual support facilitates swift, accurate, and reliable semantic annotation of the training set. This updated version of the ANALYTiC tool symbolizes a significant advancement in trajectory semantic enrichment. We hope this step will spur more innovative uses of machine learning in the domain of trajectory semantic enrichment.

# Chapter 2

# Background

## 2.1 Dimensionality Reduction Techniques

Dimensionality reduction is a fundamental technique used in machine learning and data analysis to address the challenges of high-dimensional data. It involves reducing the number of input variables or features while preserving essential information. This paper explores dimensionality reduction techniques, including Feature Selection, Feature Extraction, t-Distributed Stochastic Neighbor Embedding (t-SNE), and Principal Component Analysis (PCA).

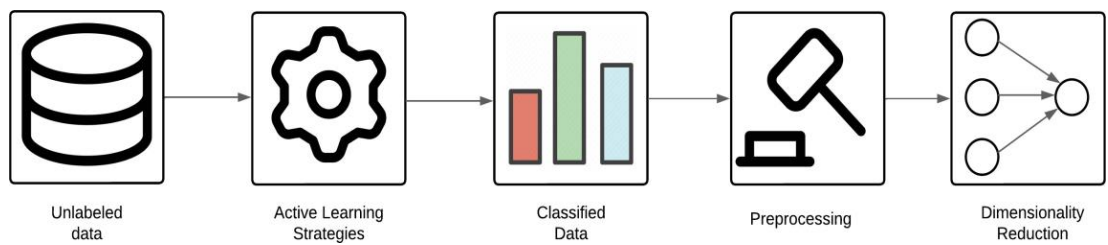

Figure 2.1: Dimensionality reduction workflow

1. Feature Selection: Feature selection aims to identify a subset of the original features that are most relevant to the problem at hand. Statistical tests, such as chi-square test or correlation analysis, assess the relationship between features and the target variable. Model-based techniques like L1 regularization (Lasso) or recursive feature elimination prioritize features based on their contribution to the model's performance [12].

2. Feature Extraction: Feature extraction transforms the original features into a lower-dimensional representation, capturing the most significant information. Principal Component Analysis (PCA) identifies orthogonal components that maximize the variance in the data. Linear Discriminant Analysis (LDA) seeks components that optimize class separability [16]. These techniques enable data simplification, improve computational efficiency, and provide insights into underlying patterns and relationships.

3. t-SNE: t-SNE is a nonlinear dimensionality reduction technique specifically designed for visualization purposes. It maps high-dimensional data points to a lower-dimensional space while preserving local structures and neighborhood relationships. By modeling pairwise similarities, t-SNE emphasizes the relative distances between data points, making it effective in revealing clusters and nonlinear structures [22].

4. PCA: is a linear dimensionality reduction technique that identifies orthogonal axes, known as principal components, capturing the highest variance in the data. These components are ranked in order of significance, with the first component explaining the most variance, the second explaining the second most, and so on. By projecting the data onto a lower-dimensional space defined by the top-k eigenvectors of the covariance matrix, PCA effectively reduces dimensionality while preserving essential information. The computation of eigenvectors and eigenvalues of the covariance matrix is central to PCA. Eigenvectors represent the principal components, while corresponding eigenvalues indicate the variance explained by each component. PCA has versatile applications in image processing, data compression, and feature extraction, particularly when the data exhibits linear structures, and interpretable results are desired. [16, 30, 36]

Dimensionality reduction techniques play a vital role in addressing the challenges of high-dimensional data. Feature selection and feature extraction methods provide ways to reduce dimensionality while retaining essential information. Furthermore, t-SNE and PCA offer powerful tools for visualizing and exploring high-dimensional data by preserving local and global structures. These techniques provide valuable insights and aid in data analysis, interpretation, and pattern recognition. The workflow of Dimensionality Reduction in ANALYTiC is summarized in Figure 2.1.

## 2.2 Decision Boundaries

In this section, we delve into the concept of decision boundaries in machine learning and its significance in classification tasks. Decision boundaries represent the boundaries that separate data points from different classes or categories. Understanding these boundaries is crucial as it provides valuable insights into classification models' behavior, performance, and generalization capabilities.

### 2.2.1 Types of Decision Boundaries

Decision boundaries can take various forms depending on the complexity and linearity of the data. In binary classification tasks, a linear decision boundary is a straight line or hyperplane that separates data points of different classes. Linear classifiers, such as Support Vector Machines (SVM) and Logistic Regression, commonly employ linear decision boundaries. However, real-world datasets often exhibit non-linearity, necessitating non-linear decision boundaries. Complex classifiers like Decision Trees, Random Forests, and Neural Networks can effectively model non-linear decision boundaries, enabling them to handle intricate data distributions [2, 15].

### 2.2.2 Visualization of Decision Boundaries

Visualizing decision boundaries is a powerful technique for better understanding a classifier's behavior. The provided Python code snippet 3.3 3.4presents an implementation of generating and visualizing decision boundaries for multiple classification models using a given dataset. The decision boundaries are visualized using contour plots and scatter plots. The mesh grid technique covers the input feature space, and the model's predictions are plotted on the mesh grid to create filled contour plots, representing the decision regions of each class. Data points from the test set are also plotted on the contour plot, enabling the assessment of the model's accuracy in separating classes [22, 35].

## 2.2.3 Significance of Decision Boundaries

Understanding decision boundaries offers several key benefits:

1. **Model Interpretability**: Decision boundaries provide an intuitive representation of how a classifier separates different classes. This transparency allows stakeholders to comprehend the reasoning behind the model's predictions, fostering trust and interpretability [20].

2. **Model Evaluation**: Decision boundaries aid in evaluating a model's performance. By visualizing how well the model separates classes, researchers can identify regions of misclassification and potential areas for model improvement [34].

3. **Model Generalization**: Decision boundaries are crucial in assessing a model's ability to generalize to new, unseen data. A well-generalized model should have decision boundaries that accurately reflect the underlying data distribution, rather than being overly influenced by the training data [15].

## 2.2.4 Practical Implications and Applications

Understanding decision boundaries has practical implications in various domains, including image recognition, natural language processing, and medical diagnosis. Researchers can leverage insights from decision boundaries to enhance model performance, interpretability, and robustness. Furthermore, decision boundary analysis can assist in identifying potential bias or unfairness in models, contributing to the development of ethically sound machine learning systems [27, 3].

The study of decision boundaries in machine learning is fundamental to the development and understanding of classification models. Researchers gain valuable insights into a model's performance and behavior by comprehending how these boundaries are derived and visualized. This understanding enables the creation of accurate and interpretable classifiers that can effectively address real-world challenges. Decision boundaries provide a critical lens through which researchers can analyze and enhance the performance of machine learning models, making them an essential component of modern research and applications in machine learning. The architecture of the Decision Boundary in ANALYTiC is shown in Figure 2.2

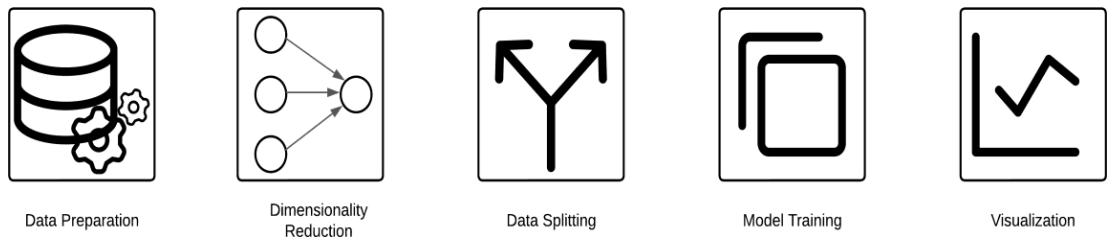

Figure 2.2: Decision surface workflow

## 2.3 Trajectory Concepts and Terminologies

A trajectory is a sequence of spatio-temporal points with associated point features like speed or direction. Point features, which can be obtained from a geolocation device or calculated from the trajectory, are static, assigned to single points, and do not change over time [13, 14]. Conversely, trajectory features are any numeric information computed using the information of the entire trajectory (e.g., average or maximal speed) [32], are dynamic and computed using the information of the entire trajectory. These features can change if the trajectory definition is modified. A semantic label or annotation is any additional semantic and/or contextual information that can be added to a trajectory [23], can be added to a trajectory to provide additional context, for example, the type of activity taking place or a specific behavior pattern.

## 2.4 Active Learning Strategies

Active Learning (AL) is a technique used when obtaining labeled data is costly or time-consuming [28]. It selects the most beneficial examples for learning, thereby reducing the required training data. A common AL scenario is pool-based sampling [19], where instances are drawn from a pool of unlabeled data based on an informativeness measure. Active learning strategies aim to choose instances for labeling that maximize the classifier's performance while respecting budget constraints.

Two popular AL strategies, uncertainty sampling (UNC) and query-by-committee (QBC) [24], alongside a random sampling (RND) baseline, are discussed. UNC labels instances where the label is most uncertain, while QBC involves a committee of models that vote on the labels of queried instances. The most informative instance is the one with the most disagreement within the committee. These AL strategies can be combined with any classifier for enhanced learning efficiency.

# Chapter 3

# Experimenting with Trajectory Data

The primary objective of this experiment is to provide an in-depth exploration and visualization of machine learning model behavior using both dimensionality reduction and decision boundary visualizations. The key goal is to transform the high-dimensional dataset into a more manageable, lower-dimensional form, while maintaining essential data characteristics. This transformation enables more efficient computational processing, easier data interpretation, and aids in mitigating the 'curse of dimensionality' that often plagues high-dimensional data. Once the data has been reduced, the decision boundaries for the machine learning models are calculated and visualized. This provides a critical view into the models' decision-making process, illustrating how the model partitions the input space to differentiate between different classes. Thus, the experiment paints a comprehensive picture of the data's structure, the efficacy of dimensionality reduction, and the decision-making paradigm of the chosen classifier, integrating these components to offer a more holistic understanding of the model's operation.

```python
# Salman-Haidri
@route('/perform_dimensionality_reduction', method='GET')
def dimensionality_reduction_endpoint():

    database = request.query.key
    file_path = database + "_label.json"

    # Read the JSON data from the file
    with open(file_path) as file:
        json_str = file.read()

    # Parse the JSON data into a dictionary
    json_data = json.loads(json_str)

    df_labels = pd.DataFrame(list(json_data.items()), columns=['id', 'label'])

    # Converting 'id' column to int
    df_labels['id'] = df_labels['id'].astype(int)

    with open('./data/' + database + ".json") as file:
        data = json.load(file)

    features = [item["line"] for item in data]

    df_all = pd.json_normalize(features)

    df_all = df_all.drop(columns=["type", "geometry.type", "geometry.coordinates"])

    # Rename columns
    df_all = df_all.rename(columns=lambda x: x.replace("properties.", ""))

    df_all = df_all.dropna()

    merged_df = df_all.merge(df_labels, left_on='tid', right_on='id')
```

Figure 3.1: Code snippet illustrating the dimensionality reduction workflow

```python
# Sort the DataFrame by 'tid'
merged_df = merged_df.sort_values('tid')
merged_df.to_csv(database + ".csv", index=False)
df = merged_df.drop(columns=['id', 'tid'])

# Split into y (label) and x (rest of the columns)
y = df['label'].values
x = df.drop(columns=['label'])
comp = (x.shape[1]) // 2
# Standardizing the data
x = StandardScaler().fit_transform(x)

pca = PCA(n_components=2)
principalComponents = pca.fit_transform(x).tolist()

pca_50 = PCA(n_components=comp)
pca_result_50 = pca_50.fit_transform(x)
tsne = TSNE(random_state=42, n_components=2, verbose=0, perplexity=40, n_iter=400).fit_transform(pca_result_50)

response_data = {
    'reduced_data': principalComponents,
    'label': y.tolist(),
    'tsne': tsne.tolist()
}
response.content_type = 'application/json'

return json.dumps(response_data)
```

Figure 3.2: Code snippet illustrating the dimensionality reduction workflow

## 3.1 Methodology for dimensionality reduction

The code[1] in Figure 3.1 and 3.2 performs the dimensionality reduction of a dataset using PCA (Principal Component Analysis) and t-SNE (t-Distributed Stochastic Neighbor Embedding). The general workflow is as follows:

1. **Load the Dataset**: This machine learning workflow starts by retrieving data from JSON files, which hold key-value pairs corresponding to unique identifiers and their respective labels. This data is translated into pandas DataFrames to facilitate efficient analysis and manipulation. During this process, columns are renamed for clarity and rows containing missing values are dropped, as these could impair the functionality of ML algorithms. Finally, the data is merged with its corresponding labels to create a comprehensive dataset, setting the stage for subsequent procedures.

2. **Data Preparation**: The DataFrame is separated into features (x) and labels (y). Feature scaling is then performed on the feature set using StandardScaler from scikit-learn, a popular ML library in Python. Feature scaling is crucial for many ML algorithms as they are sensitive to the scale of the input features. Algorithms that use a distance measure, like k-nearest neighbors (KNN), or that use gradient descent to optimize a cost function, can perform poorly if the features are not on similar scales. StandardScaler standardizes features by removing the mean and scaling to unit variance. The standard score of a sample x is calculated as:

$$z = \frac{x - \mu}{\sigma} \quad (3.1)$$

   Where $\mu$ is the mean of the training samples or zero if with_mean=False, and $\sigma$ is the standard deviation of the training samples or one if with_std=False.

3. **Dimensionality Reduction with PCA:** The Principal Component Analysis (PCA) is a dimensionality-reduction method that is often used to reduce the dimensionality of large data sets, by transforming a large set of variables into a smaller one that still contains most of the information in the large set. PCA is an unsupervised method, meaning that it does not use the label information in the data; it only considers the features. PCA works by calculating the eigenvalues and eigenvectors of the data covariance matrix. It then projects the original data along the eigenvectors corresponding to the largest eigenvalues. The resulting projected data (principal components) are uncorrelated and capture a large portion of the variance in the original data.

---

[1] https://github.com/Salman-Haidri/Honours.git

4. **Dimensionality Reduction with t-SNE:** t-Distributed Stochastic Neighbor Embedding (t-SNE) is another technique for dimensionality reduction and is particularly well suited for the visualization of high-dimensional datasets. Contrary to PCA it is not a mathematical technique but a probabilistic one. The technique maps multi-dimensional data to two or more dimensions suitable for human observation. It does this by giving each datapoint a location in a two or three-dimensional map. t-SNE uses a Student's t-distribution to compute the similarity between two points in the low-dimensional space while the similarity in the high-dimensional space is represented by a Gaussian distribution. The t-SNE algorithm then iteratively minimizes the divergence between these distributions with respect to the locations of the points in the map, effectively forcing similar data points to be modeled by nearby points and dissimilar data points to be modeled by distant points.

5. **Preparation for Visualization:** In order to visualize the transformed data, both the PCA and t-SNE transformed data, along with the corresponding labels, are sent back to the client as JSON objects. JSON, or JavaScript Object Notation, is a lightweight data-interchange format that is easy for humans to read and write and easy for machines to parse and generate. By preparing the data in this format, it can be easily sent over the network and consumed by a client-side script for visualization.

6. **Data Visualization:** On the client side, Plotly.js4 is leveraged to generate scatter plots for PCA and t-SNE transformed data. Plotly.js is a comprehensive charting library that includes over 20 chart types, such as 3D charts, statistical graphs, and SVG maps. It's particularly useful for its scatter plot capabilities, which in this case are used to display two-dimensional representations of data. Additional variables can be mapped using hue, size, and style parameters. Each data point on the scatter plot is color-coded according to its label, aiding in visualizing class distribution in the reduced feature space. This visualization can provide valuable insights into class separation and potential model classification performance.

Normalization is critical as the PCA and t-SNE transformations are sensitive to the scale of the data. StandardScaler ensures that all features contribute equally to the result, preventing high-magnitude features from dominating the transformations. Combining these dimensionality reduction techniques with a powerful visualization tool can provide significant insights into the structure of high-dimensional data and is a strong foundation for any data analysis or machine learning task.

## 3.2 Methodology for decision surface

The code snippets perform model training and makes the contour plot. The detailed workflow of it is explained below:

1. **Data Preparation:** The *Data Preparation* step starts with loading the CSV dataset into a pandas DataFrame. The labels are converted to numerical values to boost computational efficiency and compatibility. The dataset is then divided into features (X) and labels (Y), and all features are scaled using MinMaxScaler to ensure a uniform contribution to the model and mitigate bias from high-magnitude features.

2. **Dimensionality Reduction:** A significant phase in the workflow is the application of Principal Component Analysis (PCA) to the feature set (X). Dimensionality reduction to 2D is performed to enable efficient visualization and understanding of the data in a bi-dimensional space. It further helps in mitigating the curse of dimensionality, thus improving the computational performance of subsequent machine learning models.

3. **Data Splitting:** To evaluate the performance of the machine learning models, the reduced dataset is divided into a training set and a test set, ensuring an 80-20 split. This division safeguards against overfitting, thus ensuring the model's ability to generalize well to unseen data.

```python
@route('/getPlots', method='GET')
def get_plots():
    database = request.query.key
    colorScale = json.loads(request.query.colorScale)  # getting colorScale
    df = pd.read_csv(database + '.csv')

    unique_labels = df['label'].unique()
    label_mapping = {unique_labels[i]: i for i in range(len(unique_labels))}
    reverse_label_mapping = {v: k for k, v in label_mapping.items()}  # map back to original labels

    df['label'] = df['label'].map(label_mapping)

    X = df.drop('label', axis=1)
    Y = df['label']
    seed = 34
    scaler = MinMaxScaler(feature_range=(0, 1))
    X = scaler.fit_transform(X)

    names = ["Random Forest", "Extra Trees", "Gradient Boosting"]
    models = [RandomForestClassifier(random_state=seed),
              ExtraTreesClassifier(random_state=seed),
              GradientBoostingClassifier(n_estimators=10, learning_rate=.3, max_depth=1, random_state=seed)]

    pca = PCA(n_components=2)
    Xt = pca.fit_transform(X=X)
    X_train, X_test, y_train, y_test = train_test_split(Xt, Y, test_size=0.2,
                                                        random_state=seed)

    # Prepare figure
    fig2, ax2 = plt.subplots(4, 1, figsize=(6, 14))
    fig2.set_facecolor('white')
    ax2 = ax2.flatten()

    h = 0.02
    x_min, x_max = Xt[:, 0].min() - .5, Xt[:, 0].max() + .5
    y_min, y_max = Xt[:, 1].min() - .5, Xt[:, 1].max() + .5
    xx, yy = np.meshgrid(np.arange(x_min, x_max, h), np.arange(y_min, y_max, h))
```

Figure 3.3: Code snippet illustrating the decision surface workflow

4. **Decision Boundary Plot Preparation:** A grid or a meshgrid is created over the reduced 2D feature space for subsequent plotting of decision boundaries. An initial scatter plot of the reduced test data points is generated against their true classes to serve as the ground truth visualization.

5. **Model Training and Visualization:** This phase involves iterative steps for each of the three ensemble machine learning models, namely, Random Forest, Extra Trees, and Gradient Boosting. Each iteration encapsulates the following steps:

    - The respective model is trained on the training set derived from the reduced dataset. The training process involves learning the underlying patterns in the feature set to predict the labels accurately.

    - The trained model is utilized to predict the classes of the points in the

```
for i, name, model in zip(range(len(models)), names, models):
    model.fit(X_train, y_train)
    hue = model.predict(X_test)
    score = f1_score(y_test, hue, average='weighted')

    # Create a new figure for each model
    fig = Figure(figsize=(6, 3.5))
    ax = fig.subplots()

    Z = model.predict(np.c_[xx.ravel(), yy.ravel()])

    Z = Z.reshape(xx.shape)
    ax.contourf(xx, yy, Z, cmap=mcolors.ListedColormap(colorScale.values()), alpha=0.75)

    # Here we convert the predictions back to their original labels before plotting
    sns.scatterplot(x=X_test[:, 0], y=X_test[:, 1],
                    hue=pd.Series(hue).map(reverse_label_mapping), ax=ax, palette=colorScale.values())

    ax.set_title(f'{i + 1}. {name}, F-Score: {round(score, 2)}')

    # Save the figure to a PNG file
    image_file = f"images/test_{i + 1}.png"
    fig.savefig(image_file)
    plt.close(fig)
    image_paths.append(image_file)

return json.dumps({'image_paths': image_paths})
```

Figure 3.4: Code snippet illustrating the decision surface workflow

grid created in step 4. These predictions form the backbone of the decision boundary visualization.

- A contour plot is generated from the predicted classes of the points in the grid. The contour plot effectively illustrates the decision boundary of the trained model in the 2D space, thereby providing an intuitive interpretation of the model's decision-making process.

- To assess the accuracy of the model's decisions visually, the test data points are overlaid on the contour plot. The resultant plot enables a comparison between the model's predictions (decision boundaries) and the true labels of the test data.

6. **Saving and Returning the Plots:** Each decision boundary plot is diligently saved as a PNG image. The function *get_plots* finally returns the paths to these images. This provision of returning the paths rather than the images themselves enhances the modularity of the function, thereby enabling it to be easily integrated into broader workflows where the plots can be accessed and utilized as per requirement.

# Chapter 4

# Results and Discussions

## 4.1 Plots from dimensionality reduction

Upon executing the code in Figure 3.1 3.2 for dimensionality reduction, two visualizations are generated, each representing the 2-dimensional projections of the high-dimensional dataset obtained through Principal Component Analysis (PCA) and t-Distributed Stochastic Neighbor Embedding (t-SNE) respectively.In the PCA plot, the axes correspond to the first two principal components that capture the maximum variance in the data. This plot offers an insight into the data's intrinsic structure by showcasing the relationships between different data points in a reduced dimensionality. Similarly, the t-SNE plot provides a visualization of the high-dimensional data in a 2-dimensional space. However, compared to PCA, t-SNE tends to preserve the local structure of the data, making it a valuable tool for observing clusters or groups in the data. In both plots, data points are colored based on their corresponding labels. The color-coding allows for a clear understanding of the distribution of classes within the reduced feature space. The plots also facilitate easy comparison between PCA and t-SNE in terms of how well they capture the segregation of different classes in the reduced dimensions. Regarding the setup, each unique label in the dataset is mapped to a color for visualization. For instance, 'Animal' and 'Not Animal' might be mapped to 'green' and 'blue' respectively. Moreover, the dimensionality reduction techniques are fitted with the feature set X, ensuring that the visualizations embody the comprehensive patterns present in the data.

In essence, these plots 4.1 4.2 4.3 enable the visualization of high-dimensional data in a lower dimension, thereby providing a fundamental understanding of the data distribution, class segregation, and potential correlations among the data points. Both PCA and t-SNE provide valuable, albeit different, perspectives on the data. By examining these plots, one can garner insights into the structure and characteristics of the dataset, which are often not apparent in the original high-dimensional space.

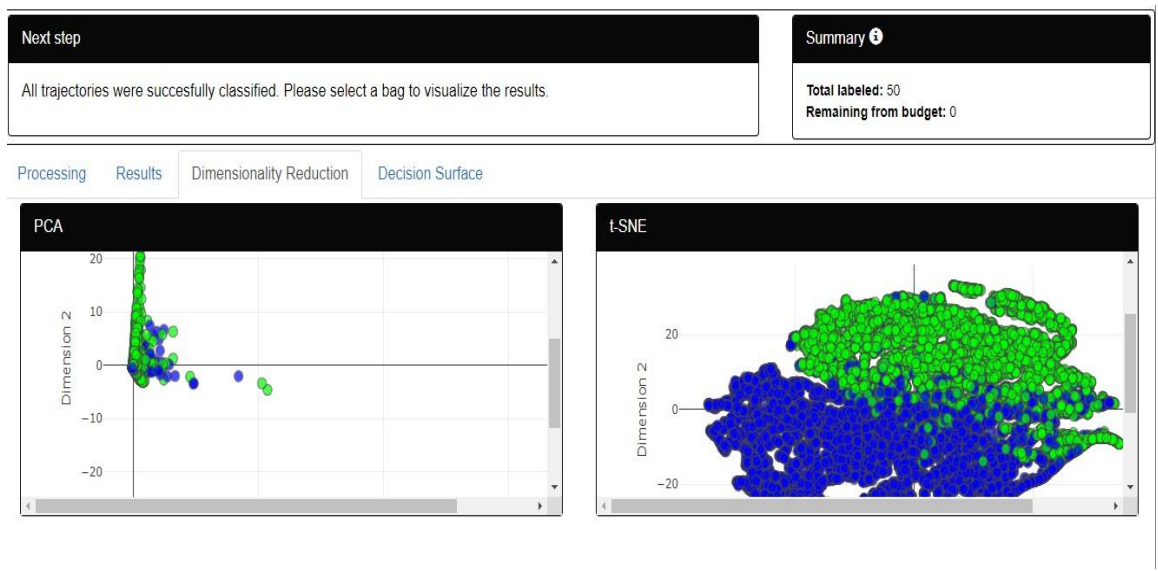

Figure 4.1: Geo life dimensionality reduction plot

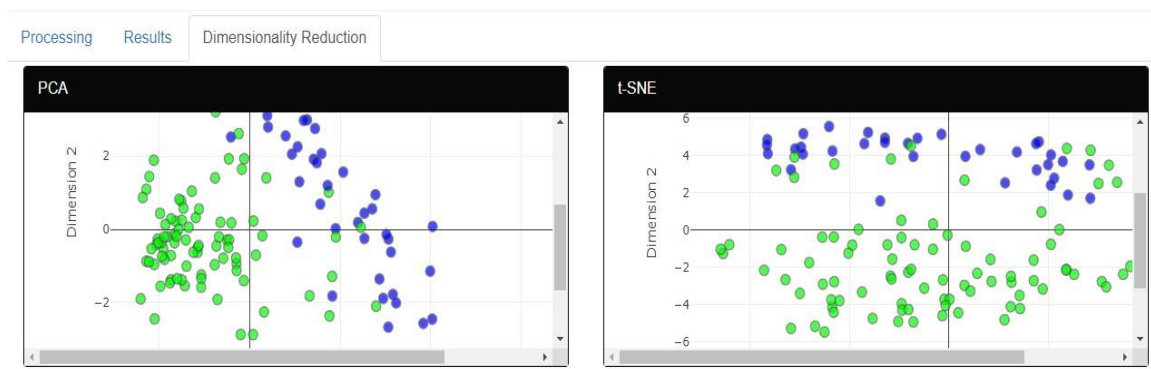

Figure 4.2: Fishing vessel dimensionality reduction plot

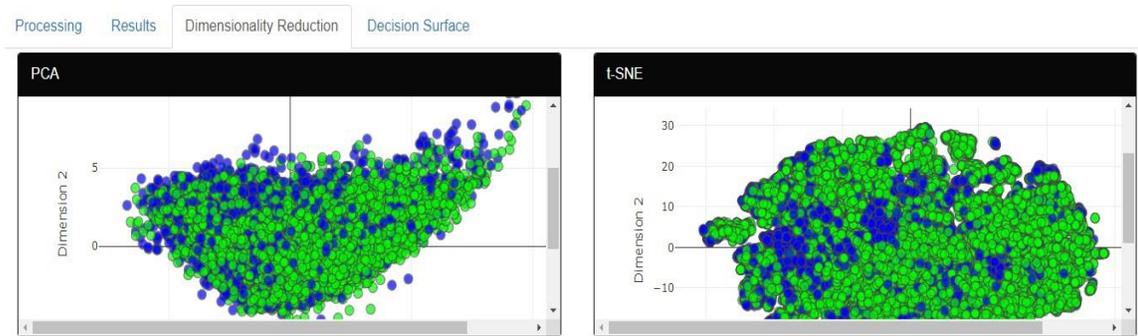

Figure 4.3: Starkey animal data dimensionality reduction plot

## 4.2 Plots from decision surface

When the provided code 3.3 3.4 is executed, an ensemble of plots are generated, each delineating a unique decision boundary or decision surface. These plots vividly depict the interface between distinct classes within the dataset, as learned by a variety of classifier models - RandomForest, ExtraTrees, and GradientBoosting.

Each plot, corresponding to a different classifier, embodies a 2-dimensional representation of the original high-dimensional data obtained after PCA transformation. The axes correspond to the two principal components encapsulating the maximum variance within the data. The data points are colored according to their actual labels, facilitating the visual comprehension of class segregation within the reduced feature space. The decision surface, demarcated by color gradients, depicts the predicted labels for all points within the plot region. The resulting colors correspond to the prediction probabilities of a particular class, enabling the visualization of the classifier's confidence across the feature space. Just like the dimensionality reductions plots, the data points are colored based on their corresponding labels.

At the setup level, each classifier, set with a fixed random seed, is trained on 80% of data and validated on 20% to ensure result reproducibility. The F-score for each classifier, a balance of precision and recall, is depicted in the plot title to indicate model performance. For visual clarity, distinct colors represent each class, such as red for 'vehicle' and blue for 'not vehicle'. The 'RdBu' color map enhances the contrast between different class boundaries on the decision surface.

The resulting plots 4.4 4.5 4.6 serve to visualize the decision boundaries and provide insight into how different classifiers partition the feature space to make predictions. Through these visualizations, one can easily compare the different classifiers based on their decision surfaces and F1 scores, and select the one that performs best.

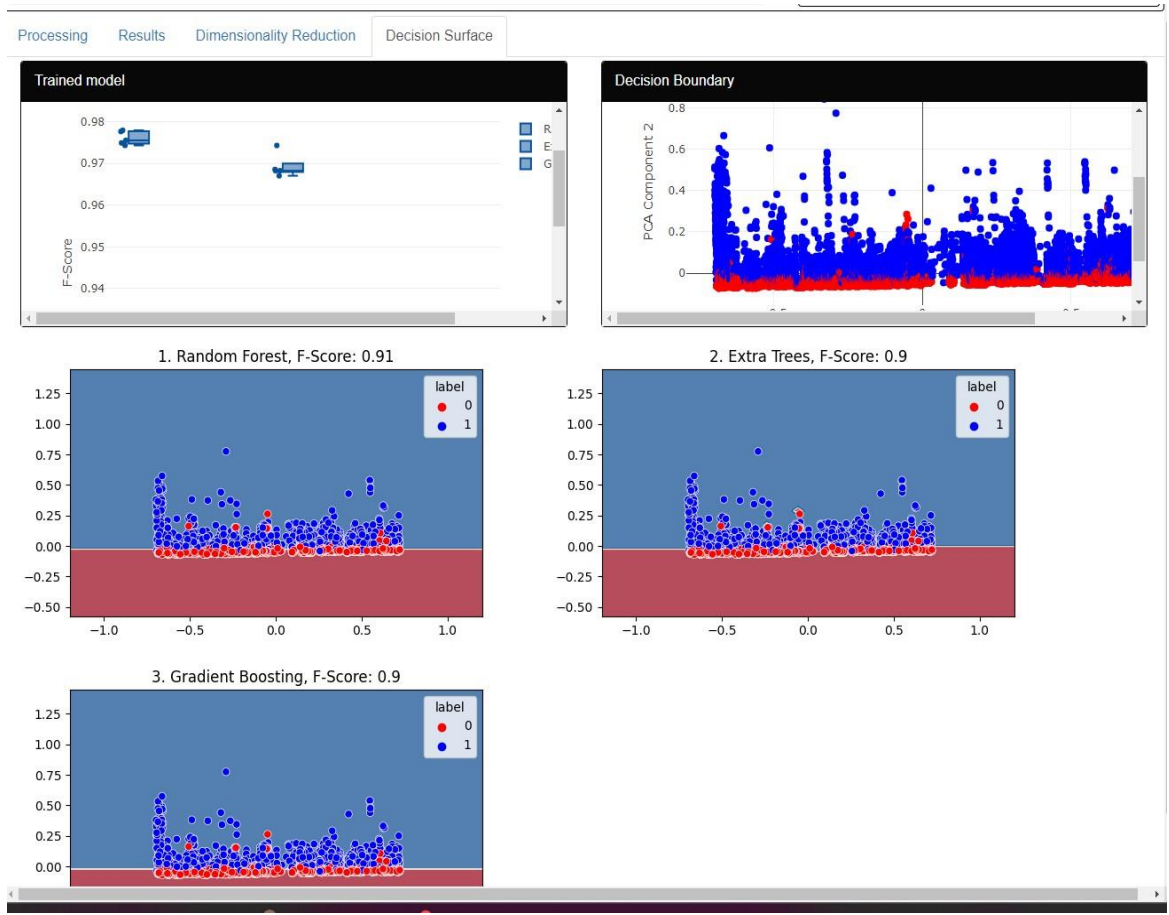

Figure 4.4: Geo life decision boundary plot

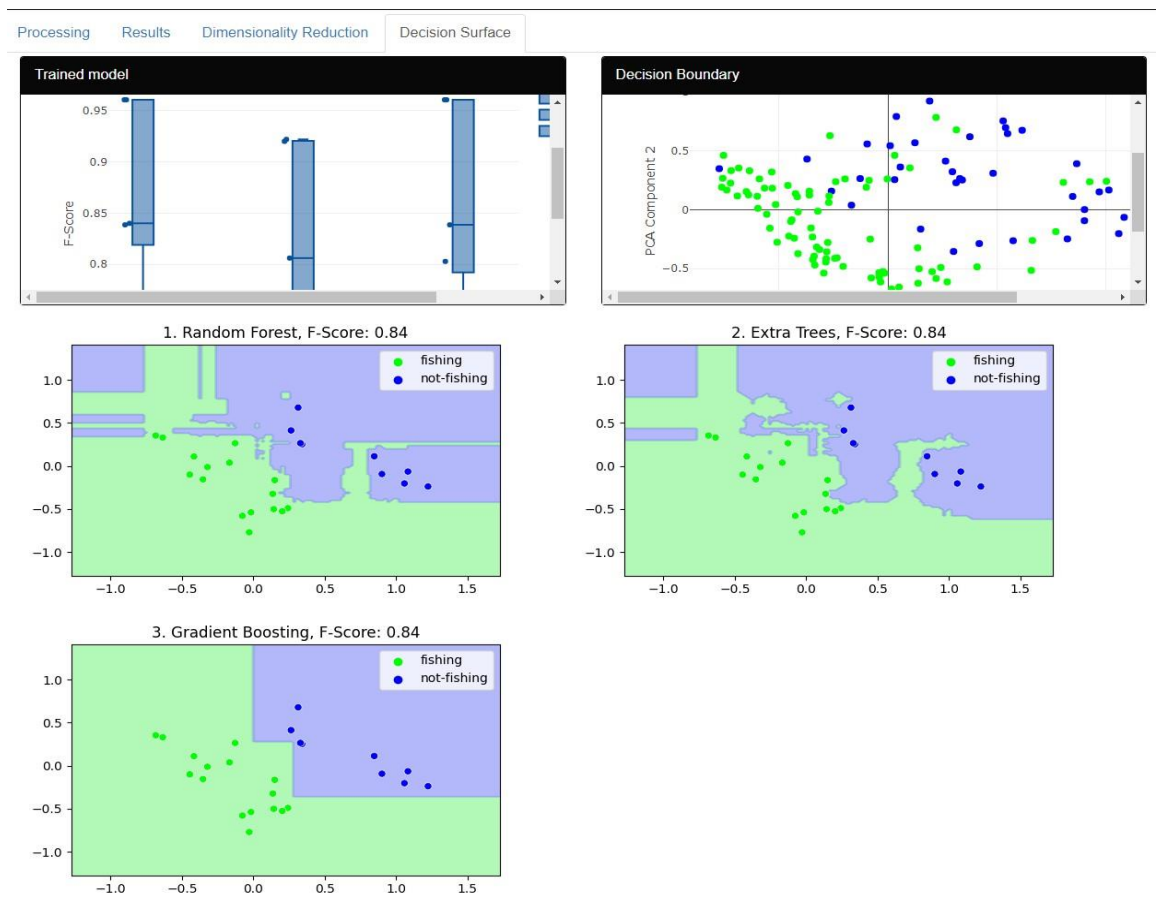

Figure 4.5: Fishing vessel decision boundary plot

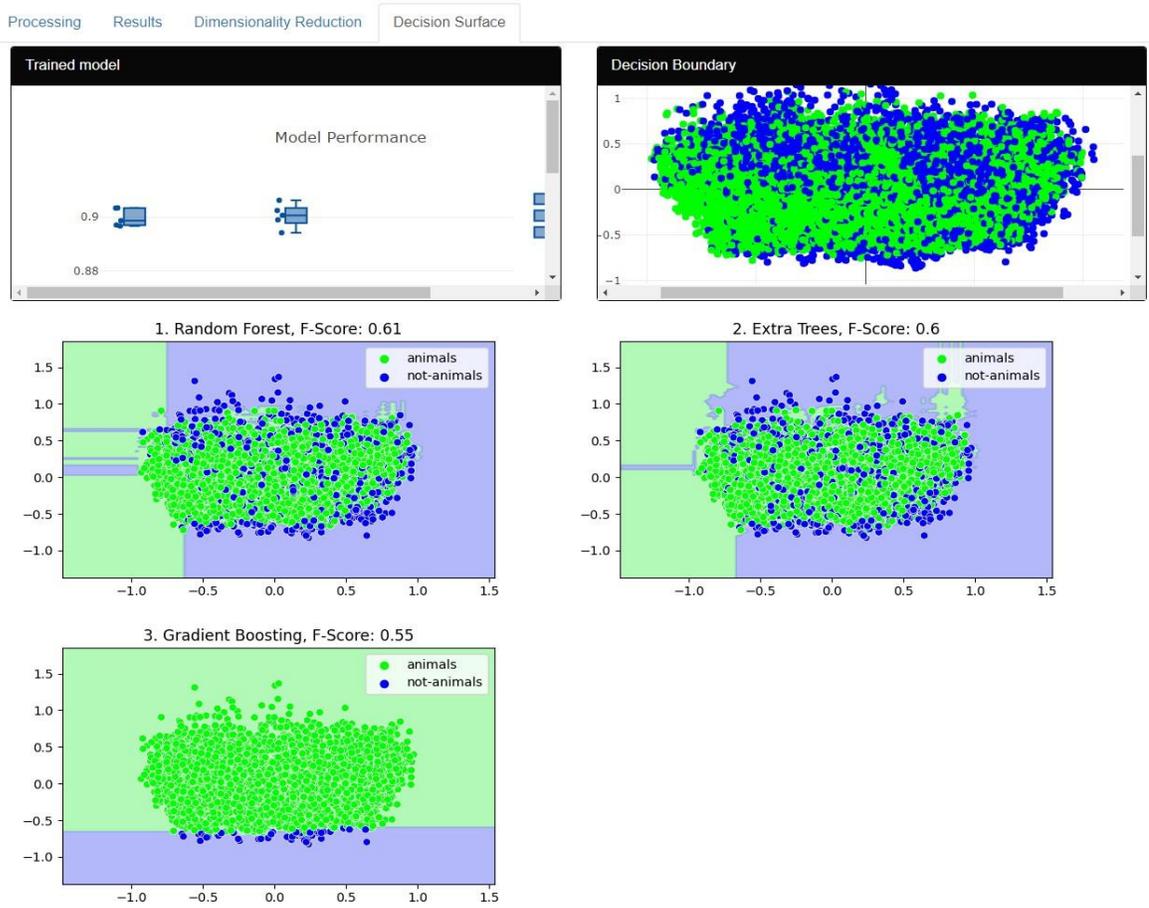

Figure 4.6: Starkey animal data decision boundary plot

# Chapter 5

# The ANALYTiC tool

The system's design consists of both front end, which provides user interaction, and back end, which manages trajectory data storage, classification algorithm execution, and active learning strategies. The architecture is shown in Figure 5.1. It is built as a web tool using HTML5 for the user interface, while user interactions are managed through JavaScript. Three integral libraries used in the project include mapbox [1] and leaflet [2] for map visualizations and interactions, and Google Charts [3] for data representation in chart format. In the latest iteration of the system, we've incorporated Plotly [4]. Plotly is a dynamic, Python-based open-source library that enables the production of interactive and high-quality visualizations, significantly enhancing the tool's data interpretation capabilities and user-friendliness.

In the original architecture of the ANALYTiC system, SOLR was used for data storage due to its fast searching capabilities. However, for the purposes of this experiment, we opted to handle the data locally within the Python environment. This choice was driven by several reasons:

Firstly, the latest Python versions provide significant improvements in handling JSON data, which is a common format for data interchange in web applications. The Python standard library includes a json module that allows reading and writing JSON data with high efficiency. Over the years, Python 3 has added substantial improvements to this module, including performance enhancements, improved fault tolerance, and ordered dictionaries [10, 11]. Specifically, Python 3's json module benefits from performance optimizations in the underlying C code that have made the json.loads() and json.dumps() functions faster than in Python 2.7. These are common functions used in handling JSON data, and their enhanced performance is particularly noticeable when working with large data sets [10]. Moreover, Python 3's json module has better fault tolerance when parsing non-standard JSON data. This makes the data handling process more robust, as it is capable of handling certain types of malformed or incorrect JSON data more effectively than the Python 2.7 version [11]. Another significant change in Python 3 is that the built-in dict class, from Python 3.7 onwards, preserves the insertion order of the keys. This means that dictionaries maintain the order in which keys were first inserted, which makes working with JSON objects in Python 3 more predictable and intuitive than in Python 2.7 [10]. Lastly, local data handling within Python enables a more fluid development process, as there's no need to manage a separate data storage service. This is advantageous for iterative testing and development, which is often necessary in a research context. In order to facilitate a smooth interface between various components of our system, we implemented the RESTful service using the bottle[5] library.

---

[1] The Mapbox Project - https://www.mapbox.com  
[2] The Leaflet Project - https://leafletjs.com  
[3] The Google Charts Project - https://developers.google.com/chart/  
[4] The Plotly Project - https://plotly.com/javascript/

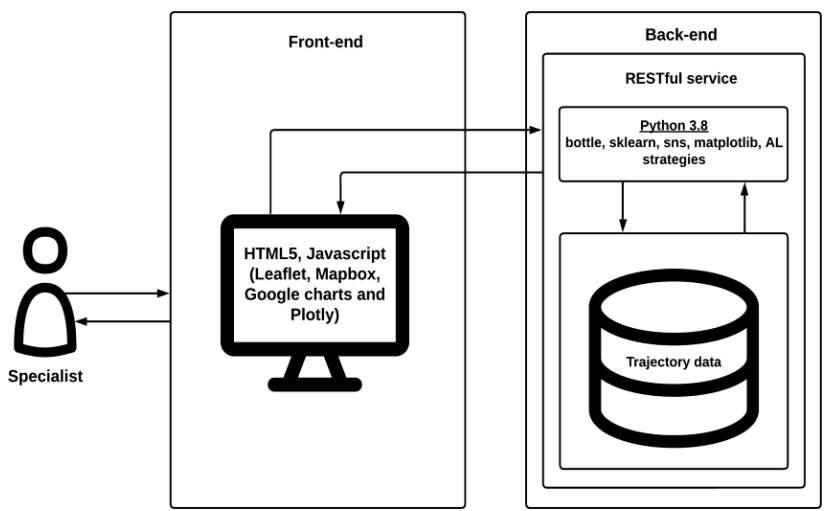

Figure 5.1: The ANALYTiC architecture

The ANALYTiC system's interface is engineered to facilitate the user experience during the trajectory labeling and active learning process. The interface, as depicted in Figure 5.2 and Figure 5.3, is composed of distinct top and bottom panels. In the central area of the interface, the user navigates through two panels: a map visualizing the trajectories and panels presenting the feature values for trajectory labeling.

---

[5]Bottle: Python Web Framework - https://bottle-rest.readthedocs.io/en/latest/

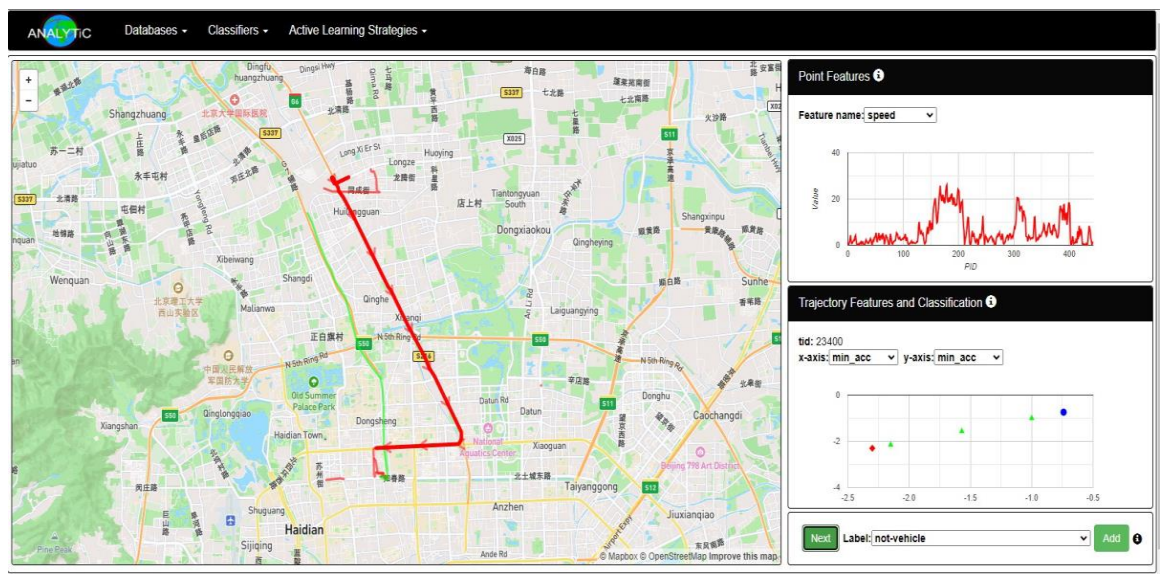

Figure 5.2: Top panel overview

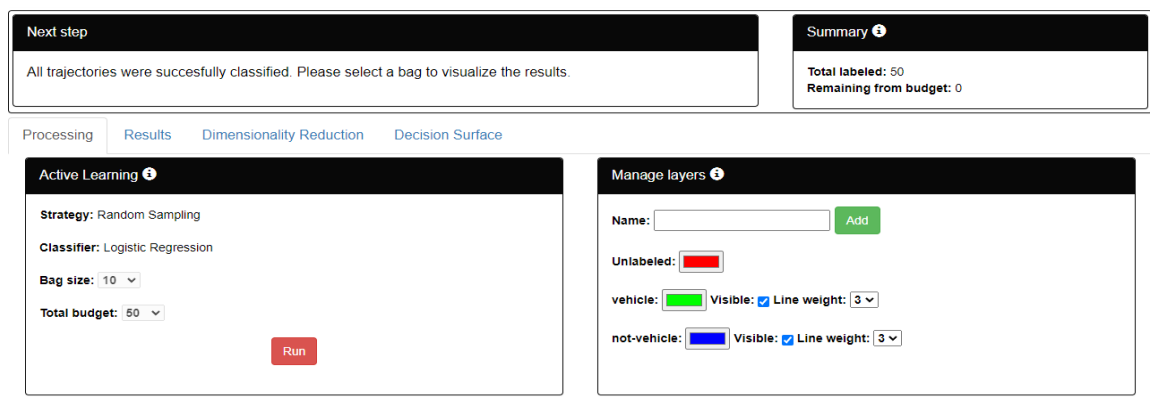

Figure 5.3: Bottom panel overview

The lower section of the interface is devoted to the active learning process, providing four panels. First, it gives an interactive guide for the user's next action. Second, it offers a concise summary of labeled and pending trajectories based on the user-selected budget. Third, it summarizes user preferences regarding the classifier,

AL strategy, bag size, and maximum budget. Lastly, it allows for label management, including addition, color selection, and visibility adjustments[32].

The active learning process is initiated by selecting the dataset, classification algorithm, and active learning strategy. After specifying the total budget of trajectories and bag size, the user is prompted to add labels and adjust visual aspects, such as color and visibility. Following this, the annotation process commences, where trajectories are selected on the map, and features like movement direction and intensity are visualized. The user is supported in the annotation process through an interactive display of numerical values of the selected features, and the process concludes when the total budget of trajectories is annotated [32].

The ANALYTiC tool enhances the user experience by merging map visualization with charts related to trajectory features and semantic labels. A noteworthy challenge is efficiently showcasing the movement and trajectory features. Therefore, two visualization solutions have been incorporated: moving arrows to represent movement direction, and color saturation to indicate feature value intensity.

Moreover, the tool offers an interactive feature where the user can connect the point features from the plots with their geographical position, aiding in understanding the behavior and location of the moving object for accurate labeling.

The user can also evaluate a trajectory chosen by the AL strategy for labeling, as well as explore correlations between trajectory features and assigned labels. After labeling several trajectories, the user can explore the trajectory feature space, contributing to a more comprehensive understanding of the object's movement patterns. This iterative process of annotating and learning enhances the user's understanding and efficiency in the active learning process [32].

Building upon the user-friendly and interactive nature of the ANALYTiC system, we are pleased to introduce an exciting new addition: the dimensionality reduction and decision surface feature, seamlessly integrated into the lower panel of the interface, alongside the results section as you can see in Figure 5.3.

This innovative feature facilitates a deeper understanding of the active learning process by visualizing the classifier's decision boundaries in a lower-dimensional space. By employing popular techniques such as Principal Component Analysis (PCA) and t-Distributed Stochastic Neighbor Embedding (t-SNE), we offer users an intuitive view of their data in reduced dimensions, aiding them in making more informed labeling decisions. This, in turn, augments the efficiency and effectiveness of the active learning process.

Upon application of PCA, each trajectory is reduced to its two principal components. As depicted in Figure 5.4, this data is then visualized in a 2D scatter plot. Each point in the plot represents a trajectory, with its position determined by the two principal components. This simplified representation allows the user to view and comprehend the structure and relationships in the data more efficiently. The platform also applies t-SNE, another dimensionality reduction technique particularly effective at preserving the local structure of the data. As illustrated in Figure 5.4, the t-SNE scatter plot provides another perspective of the data, further aiding users in their analysis and labeling tasks.

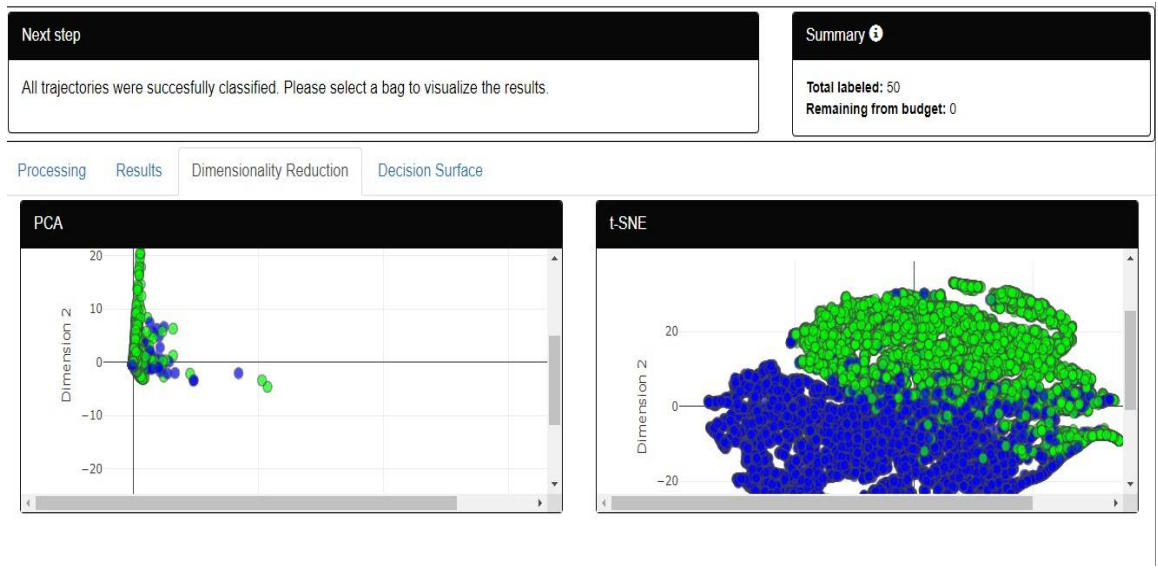

Figure 5.4: Dimensionality Reduction scatter plot

Both visualizations are interactive, with color intensities representing different trajectory features such as speed or direction variation. By studying these visualizations, users can discern patterns, correlations, and clusters within the data that may not be evident in the original high-dimensional space. For example, clusters of trajectories might correspond to specific types of behavior, allowing users to quickly and accurately assign labels.

Simultaneously, the decision surface visualization provides an insightful glimpse into the classifier's decision-making process. Users can observe the decision boundaries crafted by the various classifier and how these evolve as more trajectories are labeled, enhancing their grasp of the ongoing learning process.

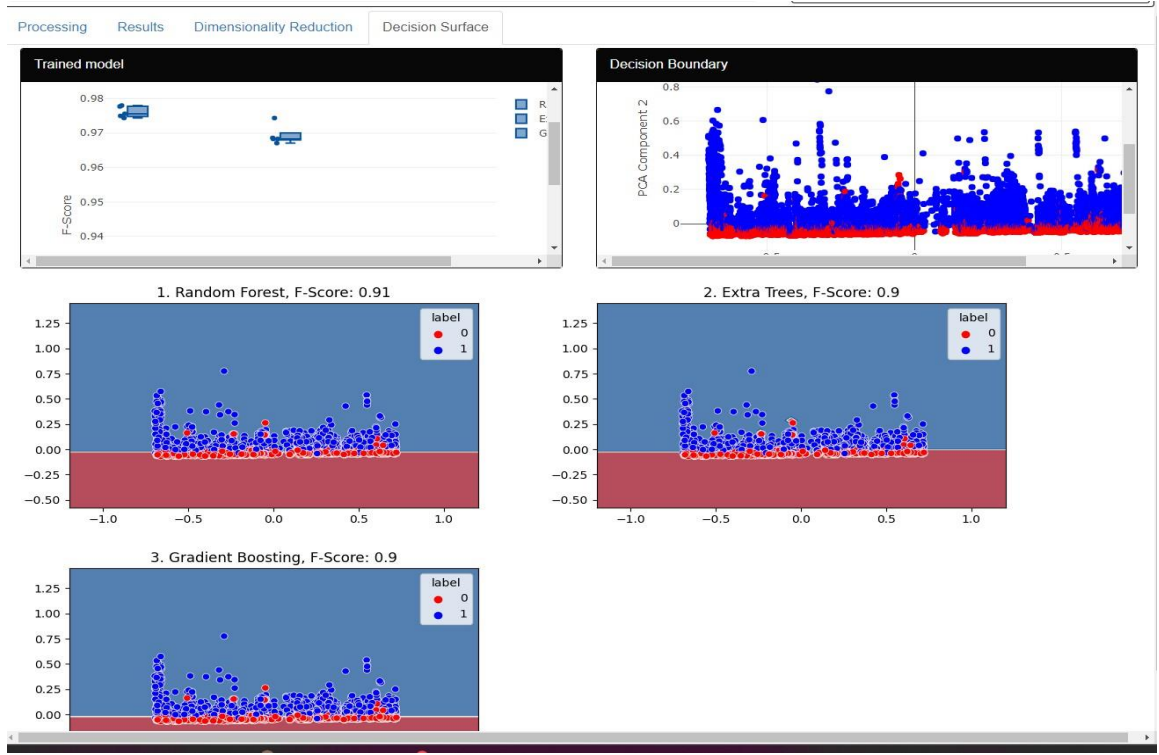

Figure 5.5: Decision surface plots

These steps culminate in a series of insightful visualizations, each corresponding to a different classifier. Each visualization, as exemplified in Figure 5.5, illustrates the decision boundaries (colored regions) and the test data points superimposed onto them. The color of the regions indicates the predicted label, while the color of the data points indicates the actual label. By observing the alignment between the colors of the regions and points, users can intuitively understand the performance of the classifier.

Moreover, the F1 score accompanying each visualization provides a quantitative measure of the classifier's performance, complementing the visual understanding.These decision boundary visualizations, coupled with the F1 scores, provide users with a comprehensive overview of the classifiers' performances. Such information can help users select the most suitable classifier for their labeling task, thereby optimizing their workflow.

These added feature fortifies the bridge between users and their active learning tasks, making the system even more powerful and comprehensible.

# Chapter 6

# Conclusion

In this paper, we delved into the realm of machine learning from an analytical perspective, focusing on understanding decision boundaries, model interpretability, and dimensionality reduction techniques. Through empirical investigations and visualization, we gained valuable insights into the behavior and performance of classification models. We explored decision boundaries, crucial for understanding classifier predictions, using contour plots and scatter plots. Model interpretability emerged as a vital aspect, enabling trust and accountability through transparent explanations. Addressing high-dimensional data, dimensionality reduction techniques like feature selection, feature extraction, t-SNE, and PCA facilitated efficient data exploration and visualization.

Our findings offer valuable guidance to researchers and practitioners in diverse domains. With a better understanding of decision boundaries and interpretability, we can build transparent and accurate machine learning models. As we progress in artificial intelligence, these analytic dimensions contribute to responsible AI applications, ensuring trust and accountability in our models for the benefit of society.